\documentclass[12pt]{iopart}

\usepackage{graphicx}
\begin{document}

\title[]{Dirac Field as a Source of the Inflation in $2+1$ Dimensional Teleparallel Gravity}

\author{Ganim Gecim and Yusuf Sucu}

\address{Department of Physics, Faculty of Science,
Akdeniz University, \\ 07058 Antalya, Turkey}
\ead{gecimganim@gmail.com and ysucu@akdeniz.edu.tr} \vspace{10pt}

\begin{abstract}
In this paper, we study early-time inflation and late-time
acceleration of the universe by non-minimally coupling the Dirac
field with torsion in the spatially flat Friedman-Robertson-Walker
(FRW) cosmological model background. The results obtained by the
Noether symmetry approach with and without a gauge term are
compared. Additionally, we compare these results with that of the
3+1 dimensional teleparallel gravity under Noether symmetry
approach. And we see that the study explains early-time inflation
and late-time acceleration of the universe.
\end{abstract}


\maketitle

\section{Introduction}

\label{sec:intro}

The recent cosmological observational data point out that our
universe has two acceleration periods: early-time and late-time
cosmic acceleration. Some cosmological difficulties, such as the
initial singularity problem, the flatness problem, and the horizon
problem, are resolved by using the early-time inflation scenario
with a scalar field as a source of quantum fluctuations \cite{Guth,
Linde}. On the other hand, the late-time inflation scenario is due
to the observed accelerating expansion of the present universe
\cite{Perll,Ries,Copeland}, where the acceleration is explained with
a negative pressure component which is called dark energy. The
models including different dark energy forms, such as cosmological
constant \cite{PeeblessRatra03,Kratochvil04}, canonical scalar field
(i.e. quintessence) \cite{CaldwellDave98,Faraoni2000}, and a phantom
field \cite{Nojiri-Odintsov03,Gibbons03}, have been discussed widely
in the context of the standard Einstein relativity.

The Teleparallel gravity is one of the modified theories which has
an alternative geometrical formulation of the Einstein's general
relativity. It was firstly introduced by Einstein to merge the
gravitation with electromagnetism \cite{Eins}. Though the spacetime
structure in the standart Einstein general relativity is described
by a Levi-Civita connection which has a free-torsion, but,
non-vanishing curvature, the spacetime structure in the teleparallel
gravity is constructed by a Weitzenbock connection which is
characterized by the metricity condition and vanishing curvature,
but it has a torsion \cite{weit}. Also, differently from the
standard Einstein relativity, the teleparallel gravity is a gauge
theory of the spacetime translation group \cite{Helh, Aldro,
Hayashi, Maluf}.

In recent decades, the 2+1 dimensional gravitation theories, such as
the standard Einstein general relativity \cite{Gidding, BTZ}, the
new massive gravity \cite{Bergs, Clement1, GY1}, the topological
massive gravity \cite{GY2, Deser, Nutku, Clement2} and the
teleparallel gravity \cite{Kawai93, Kawai94, Kawai95}, have
attracted considerable attention in the literature. In particular,
the 2+1 dimensional teleparallel theory covers an important place
among the 2+1 dimensional gravity theories because it\ has a
Newtonian limit, a black hole solution \cite{Kawai93} and a
gravitational wave solution named Siklos waves that are a special
class of the exact gravitational waves propagating on the AdS
background \cite{Blago}. Also, the other black hole solutions are
performed in the context of the 2+1 dimensional teleparallel gravity
\cite{Sousa-Maluf}, $f(T)$ gravity \cite{Ferraro-Fi}, Maxwell-$f(T)$
gravity \cite{Gonz2012} and torsion scalar coupled with a scalar
field non-minimally \cite{Gonz2014}.

As the Dirac theory has a vacuum including Zitterbewegung
oscillations between positive and negative energy states and
perfectly explains an interaction of a Dirac particle with an atomic
structure of the materials, it gives a reasonable cosmological
solutions for the early-time inflation and late-time acceleration of
the universe \cite{GYY}. Therefore, in this study, we non-minimally
couple the Dirac field with the torsion scalar and consider the
Dirac field as a source of early-time inflation and late-time
acceleration in the context of the 2+1 dimensional teleparallel
gravity on the Friedmann-Robertson-Walker (FRW) background by using
the Noether gauge symmetry approach. At same time, we show that the
cosmological solutions obtained in the context of the 2+1
dimensional teleparallel gravity are compatible with the results
which are in the coupling the Dirac field with the 3+1 dimensional
gravity theories \cite{Saha, Kremer, Y} and the 2+1 dimensional
Einstein gravity \cite{GYY}.

This paper is organized as follows. In the following section, we
find the field equations in which the Dirac field is non-minimal
coupled with the 2+1 dimensional teleparallel gravity. In Section 3,
we search the Noether symmetry for the Lagrangian which is coupled
Dirac field with the 2+1 dimensional teleparallel gravity. In
Section 4, we obtain solutions of the field equations by using
Noether symmetry approach. Furthermore, we find solutions of the
field equations under the zero-gauge term condition, and compare
these results with the results of study performed in the 3+1
dimensional teleparallel gravity. Finally, in the Section 5, we
conclude with a summary of the obtained results. Throughout the
paper, we use $c=G=\hbar =1$.

\section{The Field Equations} \label{b1}

The investigations on the interaction between a torsion and a Dirac
spinorial field have attracted attention for a long time
\cite{Datta71, Hehl73, Hehl74, Ryder-Shapiro98, Arcos04, Andrade01,
HehlHeyde76, Sabbata, And-Pere98, Maluf03}. The main purpose of
these studies are to construct a modified gravitational theories
that include a spin effect of the matter fields naturally.
Therefore, to consider the inflationary problem of the universe in
the framework of the 2+1 dimensional teleparallel gravity
non-minimally coupled with Dirac field, we start by writing the
action non-minimally coupled a Dirac field with a torsion scalar as
follows;
\begin{eqnarray}
\mathcal{A}=\int d^{3}x|e|\Bigg\{F(\Psi )T+\frac{\imath}{2}\Big[\bar{\psi}%
{\sigma}^{\mu }(x)(D_{\mu }\psi )-(D_{\mu }\bar{\psi}){\sigma}^{\mu
}(x)\psi \Big]-V(\Psi )\Bigg\},\raggedright  \label{act1}
\end{eqnarray}
where the $F(\Psi)$ and $V(\Psi )$ are the coupling functions and
the self-interaction potential, respectively. These scalar functions
depend on only function of a bilinear Dirac field
$\Psi=\bar{\psi}\psi$. And the bilinear Dirac field $\Psi$ is a
relativistic invariant quantity \cite{Griff}, where $\bar{\psi}$ is
the adjoint of the Dirac field $\psi$, which $\bar{\psi}$=$\psi
^{\dag }\sigma^{3}$. In this action, $|e|$=$\det(e_{\
\mu}^{(i)}(x))$=$\sqrt{\left\vert g\right\vert },$ where $e_{\ \mu
}^{(i)}(x)$ are dreibeins and $g$ is determinant of the metric
tensor, $g_{\mu \nu }$, defined in terms of the dreibeins and the
2+1 dimensional Minkowski spacetime metric, $\eta _{(i)(j)}$, with
the signature $(1,-1,-1)$, as $g_{\mu \nu }$=$ e_{\ \mu }^{(i)}e_{\
\nu }^{(j)}\eta_{(i)(j)}$, where $i$ and $j$ are the internal
(local) spacetime indices, and $\mu$ and $\nu$ are the external
(global) spacetime indices \cite{sucu-unal,unruh}. ${\sigma}^{\mu
}(x)$ are the Dirac matrices dependent on spacetime coordinates, and
they are defined in terms of the flat spacetime Dirac matrices,
$\bar{\sigma}^{i}$, and the dreibeins as follows;
\begin{equation}
{\sigma}^{\mu }(x)=e_{(i)}^{\ \mu }(x)\bar{\sigma}^{i}, \label{met3}
\end{equation}%
where $\bar{\sigma}^{i}$ are
\begin{equation}
\bar{\sigma}^{0}=\sigma ^{3},\ \bar{\sigma}^{1}=\imath\sigma ^{1},\ \bar{%
\sigma}^{2}=\imath\sigma ^{2}  \label{met4}
\end{equation}%
and, $\imath$ is the imaginary unit (i.e. $\imath^{2}$=$-1$),
$\sigma ^{1}$, $\sigma ^{2}$\ and $\sigma ^{3}$ are the standard
$2\times 2$ Pauli matrices \cite{sucu-unal, unver}. $T$ is the
Weitzenbock invariant (i.e. torsion scalar) defined as
\cite{Ferraro-Fi, Gonz2012};
\begin{equation}
T=S_{\rho \ \ }^{\ \mu \nu}T_{\ \mu \nu }^{\rho},\label{tscalar}
\end{equation}
where $T_{\ \mu \nu }^{\rho}$ is the Weitzenbock torsion tensor
defined in terms of the Weitzenbock connection \cite{weit,
Ferraro-Fi}, $^{W}\Gamma_{\ \mu \nu }^{\rho}$=$e_{(i)}^{\
\rho}\partial _{\nu }e_{\ \mu }^{(i)}$,
\begin{equation}
T_{\ \mu \nu }^{\rho}=e_{(i)}^{\ \rho}\left[ \partial _{\mu }e_{\
\nu }^{(i)}-\partial _{\nu }e_{\ \mu }^{(i)}\right],\label{ttensor}
\end{equation}
and $S_{\rho \ \ }^{\ \mu \nu}$ is a skew-symmetric tensor defined
as
\begin{equation}
S_{\rho \ \ }^{\ \mu \nu}=\frac{1}{4}[T_{\ \ \rho}^{\nu \mu \ }-T_{\
\ \rho}^{\mu \nu \ }+T_{\rho \ \ }^{\ \mu
\nu}]+\frac{1}{2}[\delta_{\rho}^{\mu}T_{\ \ \beta}^{\beta \nu \
}-\delta_{\rho}^{\nu}T_{\ \ \beta}^{\beta \mu \ }].\label{Stensor}
\end{equation}
Furthermore, the torsion tensor can be decomposed into three
irreducible parts in the 2+1 dimensional spacetime under the
$SO(2,1)$ group \cite{Boldo,Peixoto}: the trace part,
$t_{\mu}$=$T_{\mu \nu \ }^{\ \ \nu}$, the totally antisymmetric part
(i.e. pseudo-scalar), $\chi$=$\frac{1}{6}\varepsilon^{\mu \nu
\lambda}T_{\mu \nu \lambda}$, and the traceless symmetric tensor,
$X_{\mu \nu}$. Hence, the torsion tensor is defined in terms of
these components as follows:
\begin{equation}
T_{\mu \nu \lambda}=\chi\varepsilon_{\mu \nu
\lambda}+\frac{1}{2}(\eta_{\nu\lambda}t_{\mu}-\eta_{\mu\lambda}t_{\nu})+\varepsilon_{\mu
\nu \sigma}X_{\ \lambda}^{\sigma}\label{ttensor2}
\end{equation}
where $\varepsilon_{\mu \nu \sigma}$ is the three dimensional
Levi-Civita symbol $(\varepsilon_{012}$=$1)$. Accordingly, in a 2+1
dimensional spacetime, the minimal coupling between torsion and
Dirac field is described by the totally antisymmetric part of the
torsion tensor which is a pseudo-scalar \cite{Boldo,Peixoto}. On the
other hand, the Weitzenbock connection can be decomposed as
$^{W}\Gamma_{\ \mu \nu }^{\rho}=\Gamma_{\ \mu \nu }^{\rho}+K_{\ \mu
\nu }^{\rho}$, where $\Gamma_{\ \mu \nu }^{\rho}$ is the Christoffel
connection and $K_{\ \mu \nu }^{\rho}$ is the contorsion tensor,
$K_{\ (j)\mu }^{(i)}$=$\frac{1}{2}e_{\ \beta }^{(i)}e_{(j)}^{\ \nu
}(T_{\nu \ \ \mu }^{\ \beta }+T_{\mu \ \ \nu }^{\ \beta }-T_{\ \mu
\nu }^{\beta})$ \cite{Arcos04}. Also, the relationship between the
Weitzenbock spin connection, $\omega_{\ (j)\mu }^{(i)}$, and general
relativity (Lorentz) spin connection, $^{0}\omega_{\ (j)\mu
}^{(i)}$, becomes as $\omega_{\ (j)\mu }^{(i)}$=$^{0}\omega_{\
(j)\mu }^{(i)}+K_{\ (j)\mu }^{(i)}$. As the teleparallel gravity is
characterized by the vanishing Weitzenbock spin connection (i.e.
$\omega_{\ (j)\mu }^{(i)}$=$^{0}\omega_{\ (j)\mu }^{(i)}+K_{\ (j)\mu
}^{(i)}$=$0$) \cite{Arcos04, Andrade01, Maluf03}, the covariant
derivative of the Dirac spinor, $D_{\mu}\psi$, and its adjoint,
$D_{\mu}\bar{\psi}$, can be expressed respectively as \cite{Maluf03,
Boldo}
\begin{eqnarray}
D_{\mu }\psi =\partial _{\mu }\psi -\frac{\imath}{4}(^{0}\omega_{\
(j)\mu }^{(i)})s_{(i)}^{\ (j)}\psi, \nonumber \\
D_{\mu }\bar{\psi} =\partial _{\mu }\bar{\psi} +\frac{\imath
}{4}(^{0}\omega_{\ (j)\mu }^{(i)})s_{(i)}^{\
(j)}\bar{\psi},\label{DirrCC}
\end{eqnarray}
where the $s_{(i)}^{\ (j)}$ is a spin operator and it is defined in
terms of Dirac matrices: $s_{(i)}^{\ (j)}$=$\frac{\imath
}{2}[\overline{\sigma}_{(i)},\overline{\sigma }^{(j)}]$.
Furthermore, under these considerations, the minimal coupling the
Dirac spinor to the torsion become equivalent to the that of
curvature. Accordingly, the Dirac spinor can be consistently coupled
to torsion as indicated in previously works \cite{Maluf03, Formiga,
Mielke, Obuk}.

The variation of the action (\ref{act1}) with respect to
$\bar{\psi}$ and $\psi$ gives us the Dirac equation and its adjoint
as follows:
\begin{eqnarray}
&& \imath\sigma^{\mu}(x)D_{\mu }\psi +\left[F^{\prime}T-V^{\prime
}\right] \psi =0 \label{DirrC1} \\
&&\imath(D_{\mu }\overline{\psi})\sigma^{\mu}(x)-\left[
F^{\prime}T-V^{\prime}\right] \overline{\psi }=0 \label{DirrC2}
\end{eqnarray}
respectively, where the prime denotes the derivative with respect to
the bilinear $\Psi$. Additionally, the variation of the action with
respect to the dreibein fields, $e_{\ \sigma}^{(i)}(x)$, yields the
equation of motion as follows;
\begin{eqnarray}
F\left[Te_{(i)}^{\ \sigma}-4T_{\ \nu (i)}^{\alpha }S_{\alpha \ \
}^{\ \nu \sigma}+4e^{-1}\partial _{\alpha }(eS_{(i) \ \ }^{\ \alpha
\sigma })\right]+4F^{\prime}(\partial _{\alpha }\Psi )S_{(i) \ \
}^{\ \alpha \sigma }=\mathrm{T}_{(i)}^{\ \sigma },\label{motion1}
\end{eqnarray}
where $T_{\ \nu (i)}^{\alpha }$=$e_{(i)}^{\ \lambda }T_{\ \nu
\lambda }^{\alpha }$, $S_{(i) \ \ }^{\ \alpha \sigma }$=$e_{(i)}^{\
\lambda}S_{\lambda \ \ }^{\ \alpha \sigma }$ and the
$\mathrm{T}_{(i)}^{\ \sigma }$ is the modified energy-momentum
tensor of the Dirac field and its explicit form is expressed as
\begin{eqnarray}
\mathrm{T}_{\beta \mu}=\frac{\imath}{2}\left[ \overline{\psi
}\sigma_{\beta}(x)D_{\mu}\psi -\left(D_{\mu}\overline{\psi}\right)
\sigma_{\beta}(x)\psi \right]+g_{\beta \mu}\left[ \left(
F^{\prime}T-V^{\prime}\right)\Psi +V\right].\label{tortens}
\end{eqnarray}

To investigate the inflationary problem in the context of the 2+1
dimensional teleparallel gravity, we will consider the 2+1
dimensional FRW spacetime background that is spatially flat,
homogeneous and isotropic universe,
\begin{equation}
ds^{2}=dt^{2}-a^{2}(t)[dx^{2}+dy^{2}],  \label{FRW}
\end{equation}%
where $a(t)$ is the scale factor of the universe. For the metric,
the dreibein fields and their dual are expressed as $e_{\ \mu
}^{(i)}$=$diag(1,a(t),a(t))$ and $e_{(i)}^{\ \mu
}$=$diag(1,1/a(t),1/a(t))$, respectively. Accordingly, the non-null
components of the torsion tensor (\ref{ttensor}) and the $S_{\rho \
\ }^{\ \mu \nu}$ skew-symmetric tensor (\ref{Stensor}) are given by,
respectively,
\begin{eqnarray}
T_{\ 0 1}^{1}=T_{\ 0 2}^{2}=H, \ \ \ S_{1 \ \ }^{\ 1 0}=S_{2 \ \
}^{\ 2 0}=\frac{H}{2},\label{sclartors}
\end{eqnarray}
where $H$=$\frac{\dot{a}}{a}$ is the Hubble parameter and the dot
represents differentiation with respect to cosmic time, $t$.
Moreover, using the equation (\ref{tscalar}), the torsion scalar
becomes $T$=$-2H^{2}$. Hence, the Dirac equation (\ref{DirrC1}) and
its adjoint (\ref{DirrC2}) are reduced as, respectively,
\begin{eqnarray}
&&\dot{\psi}+H\psi +\imath V^{\prime }\sigma ^{3}\psi +2\imath
H^{2}F^{\prime}\sigma^{3}\psi =0,  \label{dirc1} \\
&&\dot{\bar{\psi}}+H\bar{\psi}-\imath V^{\prime }\bar{\psi}\sigma
^{3}-2\imath H^{2}F^{\prime }\bar{\psi}\sigma ^{3}=0.  \label{dirc2}
\end{eqnarray}
Furthermore, the $"00"$ and $"11"$ components of the equation of
motion (\ref{motion1}) give us the following:
\begin{equation}
H^{2}=\frac{V}{2F}  \label{fried1}
\end{equation}
and
\begin{equation}
\frac{\ddot{a}}{a}=-\frac{2HF^{\prime }\dot{\Psi}+\left[
2H^{2}F^{\prime }+V^{\prime }\right] \Psi -V}{2F}  \label{fried2}
\end{equation}
respectively. Also, the $"22"$ component is equivalent to the $"11"$
component.

In order to solve the field equations (\ref{fried1}) and
(\ref{fried2}), a suitable forms of the unknown functions $F(\Psi)$
and $V(\Psi)$ must be given. However, in this study, we calculate
the forms of these functions by using the Noether gauge symmetry
approach method. For this reason, firstly, we must construct the
Lagrangian of the system defined by equation (\ref{act1}).
Therefore, with similar calculations in the \cite{GYY}, the
point-like Lagrangian of the system can be written from the action
(\ref{act1}) in the following form:
\begin{eqnarray}  \label{lag}
L=2F\dot{a}^{2}-\frac{\imath a^{2}}{2}\left( \bar{\psi}\sigma ^{3}\dot{\psi}-%
\dot{\bar{\psi}}\sigma ^{3}\psi \right) +a^{2}V.
\end{eqnarray}
The Dirac's equations for the spinor field $\psi $ and its adjoint
$\bar{\psi}$ are obtained from the Lagrangian (\ref{lag}) such that
the Euler-Lagrange equations of the $\psi $ and $\bar{\psi}$ are,
respectively,
\begin{eqnarray}
&&\dot{\psi}+H\psi +\imath V^{\prime }\sigma ^{3}\psi
+2\imath H^{2}F^{\prime }\sigma^{3}\psi =0,  \label{dirac1} \\
&&\dot{\bar{\psi}}+H\bar{\psi}-\imath V^{\prime }\bar{\psi}\sigma
^{3}-2\imath H^{2}F^{\prime }\bar{\psi}\sigma ^{3}=0, \label{dirac2}
\end{eqnarray}%
Furthermore, letting the point-like Lagrangian (\ref{lag}) and
Dirac's equations, we find the second order Euler-Lagrange equation
for $a$, i.e. the acceleration equation,
\begin{equation}
\frac{\ddot{a}}{a}=-\frac{p_{_{f}}}{2F},  \label{acce}
\end{equation}%
where $p_{_{f}},$
\begin{eqnarray}  \label{pressure}
p_{_{f}}=2HF^{\prime }\dot{\Psi}+\left[ 2H^{2}F^{\prime }+V^{\prime
}\right] \Psi -V,
\end{eqnarray}
is the pressure of the Dirac field. Furthermore, the constraint
equation for the energy function ($E_{L}=0$) associated with the
point-like Lagrangian (\ref{lag}) is written as
\begin{equation}
E_{L}=\frac{\partial {L}}{\partial {\dot{a}}}\dot{a}+\frac{\partial {L}}{%
\partial {\dot{\psi}}}\dot{\psi}+{\dot{\bar{\psi}}}\frac{\partial {L}}{%
\partial {\dot{\bar{\psi}}}}-L,  \label{hamilton}
\end{equation}%
\cite{Demi, Cap1}, which it is equivalent to the Friedmann equation:
\begin{equation}
H^{2}=\frac{\rho _{_{f}}}{2F},  \label{fried}
\end{equation}%
where $\rho _{_{f}}$ represents the effective energy density and it
is given by the self-interaction potential of a Dirac field in the
following form;
\begin{equation}
\rho _{_{f}}=V.  \label{ener}
\end{equation}
As can be seen, the Dirac equation (\ref{dirc1}) and its adjoint
(\ref{dirc2}) as well as the equations of motion (\ref{fried1}) and
(\ref{fried2}) were recovered again by using the Lagrangian
formalism.


\section{The Noether symmetry approach}

\label{ns}

The symmetry conception has always played a central role in physics
because it is directly associated with the conservation laws of a
dynamical system \cite{Arnold}. These connections between symmetries
and conservation laws of a dynamical system is expressed by the
Noether's theorem \cite{noo}. So, in this study, the existence of
Noether symmetries lead to a specific form of coupling function and
the self-interaction potential, which they are important for
obtaining the exact solutions of the fields equations \cite{Demi,
Cap1, Wei, hussain11, Pali1, Pali3, Pali4}. Mathematically, the
Noether symmetry condition for the system where the Dirac field
coupled with the 2+1 dimensional gravity, with a gauge term, $B$,
can express as follows \cite{GYY, Bluman, Stephani, hussain11,
Pali2};
\begin{equation}
\mathbf{X}^{[1]}L+L(D_{t}\tau)=D_{t}B,  \label{noether}
\end{equation}
where $B$=$B(t,a,\psi _{j},\psi _{j}^{\dagger },\dot{a},\dot{\psi
_{j}},\dot{\psi _{j}^{\dagger }})$ is a gauge term, $D_{t}$ is the
operator of the total differentiation with respect to $t$
\begin{equation}
{D_{t}}=\frac{\partial }{\partial t}+\dot{a}\frac{\partial }{\partial a}%
+\sum_{j=1}^{2}\left( \dot{\psi _{j}}\frac{\partial }{\partial \psi _{j}}+%
\dot{\psi _{j}^{\dagger }}\frac{\partial }{\partial \psi _{j}^{\dagger }}%
\right),\label{totd}
\end{equation}
and $\mathbf{X}^{[1]}$, where
\begin{eqnarray}
\mathbf{X}^{[1]}&= \mathbf{X}+\left(D_{t}\alpha -\dot{a}D_{t}\tau\right)\frac{\partial }{\partial \dot{a}}%
+\sum_{j=1}^{2}\left[\left(D_{t}\beta _{j}-\dot{\psi
_{j}}D_{t}\tau\right)\frac{\partial }{\partial
\dot{\psi_{j}}}\right]\nonumber\\&+\sum_{j=1}^{2}\left[\left(D_{t}\gamma
_{j}-\dot{\psi _{j}^{\dagger }}D_{t}\tau\right)\frac{\partial
}{\partial \dot{\psi _{j}^{\dagger }}}\right],\label{fpro}
\end{eqnarray}
is the first-order prolongation of the vector field $\mathbf{X}$
given by
\begin{eqnarray}
\mathbf{X}=\tau \frac{\partial }{\partial t}+\alpha \frac{\partial }{%
\partial a}+\sum_{j=1}^{2}\left( \beta _{j}\frac{\partial }{\partial \psi
_{j}}+\gamma _{j}\frac{\partial }{\partial \psi _{j}^{\dagger
}}\right),\label{vecf}
\end{eqnarray}
where the coefficients $\tau$, $\alpha$, $\beta_{j}$ and
$\gamma_{j}$ are dependent on the variables $t,a,\psi _{j},\psi
_{j}^{\dagger },\dot{a},\dot{\psi _{j}},\dot{\psi _{j}^{\dagger }}$.
Letting the spinor field $\psi =(\psi _{1},\psi _{2})^{T}$ and its
adjoint, $\bar{\psi}=\psi ^{\dag }\sigma ^{3}$, the point-like
Lagrangian (\ref{lag}) is reduced to the following form:
\begin{equation}
L=2F\dot{a}^{2}-\frac{\imath a^{2}}{2}\left[ \sum_{j}^{2}(\psi _{j}^{\dagger }\dot{%
\psi _{j}}-\dot{\psi _{j}^{\dagger }}\psi _{j})\right] +a^{2}V.
\label{la2}
\end{equation}%
Hence, the point-like Lagrangian (\ref{la2}) and the Noether gauge
symmetry condition (\ref{noether}) leads to the following partial
differential
equations obtained from the fact that the coefficients of the $\dot{a}%
^{3},\dot{a}^{2},\dot{a},\dot{\psi _{j}},\dot{\psi _{j}^{\dagger }},\dot{a}%
\dot{\psi _{j}},\dot{a}\dot{\psi _{j}^{\dagger }},\dot{a}^{2}\dot{\psi _{j}}$%
, and $\dot{a}\dot{\psi _{j}^{\dagger }}$ vanish, separately:
\begin{equation*}
2 F \left(2 \frac{\partial \alpha}{\partial a}-\frac{\partial
\tau}{\partial t}\right) +2 F^{\prime }\sum_{j=1}^2
\epsilon_{j}\left(\beta_{j}\psi_{j}^\dagger +
\gamma_{j}\psi_{j}\right)= 0,  \label{neq1}
\end{equation*}
\begin{eqnarray*}
4 F \frac{\partial{\alpha}}{\partial{\psi_{j}}} =0,\quad\quad 4 F
\frac{\partial{\alpha}}{\partial{\psi_{j}^\dagger}} =0, \label{neq2}
\end{eqnarray*}
\begin{eqnarray*}
2 F \frac{\partial{\tau}}{\partial{\psi_{j}}} =0,\quad\quad 2 F
\frac{\partial{\tau}}{\partial{\psi_{j}^\dagger}} =0, \qquad 2 F
\frac{\partial{\tau}}{\partial{a}}=0, \label{neq22}
\end{eqnarray*}
\begin{eqnarray}
\imath \alpha \psi_{j} + \frac{\imath a}{2} \beta_{j} - \frac{\imath
a}{2} \sum_{i=1}^2\left(\frac{\partial{\beta_{i}
}}{\partial{\psi_{j}^\dagger}}\psi_{i}^\dagger - \frac{\partial{\gamma_{i}}}{%
\partial{\psi_{j}^\dagger}}\psi_{i}\right)+ a V
\frac{\partial{\tau}}{\partial{\psi_{j}^\dagger}} -
\frac{1}{a}\frac{\partial{B}}{\partial{\psi_{j}^\dagger}} = 0,
\label{neq3}
\end{eqnarray}
\begin{eqnarray*}
\imath \alpha \psi_{j}^\dagger + \frac{\imath a}{2} \gamma_{j} +
\frac{\imath a}{2} \sum_{i=1}^2\left(\frac{\partial{\beta_{i}
}}{\partial{\psi_{j}}}\psi_{i}^\dagger -
\frac{\partial{\gamma_{i}}}{\partial{\psi_{j}}}\psi_{i}\right)- a V
\frac{\partial{\tau}}{\partial{\psi_{j}}}
+\frac{1}{a}\frac{\partial{B}}{\partial{\psi_{j}}}=0,  \label{neq4}
\end{eqnarray*}
\begin{eqnarray*}
4F \frac{\partial {\alpha}}{\partial {t}}-\frac{\imath a^2}{2}
\sum_{j=1}^2\left(\frac{\partial{\beta_{j}
}}{\partial{a}}\psi_{j}^\dagger - \frac{\partial{\gamma_{j}}}{
\partial{a}}\psi_{j}\right)+ a^2 V
\frac{\partial {\tau}}{\partial {a}}-\frac{\partial {B}}{\partial
{a}}=0 \label{neq7}
\end{eqnarray*}
and also, the remaining expressions satisfy the following equation:
\begin{eqnarray}
(2 \alpha+a \frac{\partial{\tau}}{\partial {t}}) V
-\frac{1}{a}\frac{\partial{B}}{\partial {t}} + a V^{\prime
}\sum_{j=1}^2 \epsilon_{j} \left(\beta_{j}\psi_{j}^\dagger +
\gamma_{j}\psi_{j}\right)\nonumber \\-\frac{\imath
a}{2}\sum_{j=1}^2\left(\psi_{j}^\dagger
\frac{\partial{\beta_{j}}}{\partial{t}}-\psi_{j}
\frac{\partial{\gamma_{j}}}{\partial{t}}\right) = 0, \label{neq9}
\end{eqnarray}
where
\begin{eqnarray}
\epsilon_{j}=\left\{
\begin{array}{c}
1~~ ~~\textstyle{for}~~ j=1\;\cr -1~~ \textstyle{for}~~ j=2.\;%
\end{array}
\right.\label{epsil}
\end{eqnarray}
Then, the complete solutions of the (\ref{neq3}) are obtained as
follows:
\begin{eqnarray}
\alpha &=-\frac{c_{1}(k+1)}{2(k-1)}a, \quad
\beta_{j}=\frac{c_{1}(k+1)}{2(k-1)}\psi _{j}+\epsilon _{j}\beta
_{0}\psi_{j},  \nonumber \\&
\gamma_{j}=\frac{c_{1}(k+1)}{2(k-1)}\psi _{j}^{\dagger }-\epsilon
_{j}\beta _{0}\psi _{j}^{\dagger },\quad \tau = c_{1}t+c_{2},\quad
B=c_{4} \label{vec1}
\end{eqnarray}
and the coupling function, $F(\Psi)$, is found as
\begin{equation}
F(\Psi)=f_{0}\Psi^{\frac{2 k}{k+1}},  \label{coupling}
\end{equation}
where the $c_{1}$, $c_{2}$, $c_{4}$, $f_{0}$ and $k$ ($k\neq 1$) are
integration constants. Finally, inserting (\ref{vec1}) and
(\ref{coupling}) in (\ref{neq9}), the self-interaction potential is
determined as
\begin{eqnarray}
V(\Psi) = \lambda\Psi^{\frac{2}{k+1}},  \label{pot1}
\end{eqnarray}
where $\lambda $ is integration constants.

Using (\ref{vec1}), the corresponding Noether gauge symmetry
generators are obtained as,
\begin{eqnarray*}
\mathbf{X}_{0} & = & \frac{\partial}{\partial{t}},
\end{eqnarray*}
\begin{eqnarray}
\mathbf{X}_{1} & = & t\frac{\partial}{\partial{t}} - \frac{k+1}{2(k-1)}\left[a%
\frac{\partial}{\partial{a}}-\sum_{i=1}^2(\psi_{i}\frac{\partial}{\partial{%
\psi_{i}}}+ \psi_{i}^{\dagger}\frac{\partial}{\partial{\psi_{i}}^{\dagger}})%
\right],  \label{gen2}
\end{eqnarray}
\begin{eqnarray*}
\mathbf{X}_{2}& = &\sum_{i=1}^2\epsilon_{i}(\psi_{i}\frac{\partial}{\partial{%
\psi_{i}}}-\psi_{i}^{\dagger}\frac{\partial}{\partial{\psi_{i}}^{\dagger}}).
\end{eqnarray*}
Furthermore, these generators satisfy the following commutation
relations;
\begin{eqnarray}  \label{commu1}
[\mathbf{X}_{0},\mathbf{X}_{1}] = \mathbf{X}_{0}, \quad [\mathbf{X}_{0},%
\mathbf{X}_{2}] =[\mathbf{X}_{1},\mathbf{X}_{2}] =0
\end{eqnarray}

Thanks to the Noether theorem, if the vector field $\mathbf{X}$ is a
Noether gauge symmetry corresponding to the Lagrangian $L$, then,
the following equivalent is a first integral (i.e. conserved
quantities of the system) associated with $\mathbf{X}$:
\begin{eqnarray}
\raggedright{I}&=\tau L + \left(\alpha-\tau \dot{a}\right)
\frac{\partial L}{\partial \dot{a}}+ \sum_{j=1}^{2}\left[
(\beta_{j}-\tau \dot{\psi_{j}}) \frac{\partial L}{\partial
\dot{\psi_{j}}}\right]\nonumber
\\& +\sum_{j=1}^{2}\left[(\gamma_{j}-\tau \dot{\psi_{j}^\dagger})
\frac{\partial L}{\partial \dot{\psi_{j}^\dagger}}\right]- B.
\label{Frsti}
\end{eqnarray}
Moreover, there are three first integrals (or conserved quantities)
associated with the Noether gauge symmetries:
\begin{eqnarray}  \label{frstI-1}
I_{0} &=& -2 F \dot{a}^2+a^2 V,\\
I_{1} &=& tI_{0}-\frac{2F(k+1)}{(k-1)}a\dot{a},\label{frstI-2}\\
I_{2} &=& -\frac{\imath a^2 \Psi}{2}.\label{frstI-3}
\end{eqnarray}
It is important to note that the first integral (\ref{frstI-1}) is
associated to the energy function (\ref{hamilton}), since the first
integral $I_{0}$ vanishes identically.

\section{The solutions of the field equations}

\label{fieldeqns}

To determine the time evolution of the scale factor, $a(t)$,
firstly, we need to know the form of bilinear function, $\Psi$.
Thus, we insert the coupling function $F$ given by (\ref{coupling})
in the Dirac equation (\ref{dirac1}) and in its adjoint
(\ref{dirac2}) we gets
\begin{eqnarray}
\dot{\Psi} + 2 \frac{\dot{a}}{a}\Psi=0.  \label{sol}
\end{eqnarray}
Then, the form of the bilinear function is calculated as
\begin{eqnarray}
\Psi = \frac{\Psi_0}{a^2},  \label{sol2}
\end{eqnarray}
where $\Psi _{0}$ is a constant of integration. Then, the first
integral equation (\ref{frstI-3}) is found $I_{2}=-\frac{\imath \Psi
_{0}}{2}$. Furthermore, inserting the equations (\ref{sol2}) and the
coupling function (\ref{coupling}) in the first integral
(\ref{frstI-2}), we gets
\begin{eqnarray}
\dot{a}+K a^\frac{3k-1}{k+1} = 0,  \label{frstI-22}
\end{eqnarray}
where
\begin{eqnarray}
K=\frac{(k-1)I_{1}}{2(k+1)f_{0}\Psi_{0}^\frac{2k}{k+1}},
\label{const1}
\end{eqnarray}
and then integrating this equation, the scale factor $a(t)$ finds as
\begin{eqnarray}
a(t) = \left[\frac{2 K (k-1)}{k+1} t
+a_{0}\right]^{\frac{k+1}{2(1-k)}}, \label{scale1}
\end{eqnarray}
where $a_{0}$ is an integration constant. Using the solution
(\ref{scale1}) in the acceleration equation (\ref{acce}) and in the
Friedmann equation (\ref{fried}), we obtain a constraint relations
between the constants as $\lambda
$=$I_{1}^{2}(k-1)^{2}/2\Psi_{0}^{2}f_{0}(k+1)^{2}$. Under these
conditions, the scale factor that is obtained from the Noether gauge
symmetry represents a power law expansion for the universe.
Moreover, using the definition of the deceleration parameter $q$,
namely $q$=$-a\ddot{a}/\dot{a}^{2}$, it can be seen that the
evolution of the universe has undergone three different processes.
Using (\ref{scale1}), the deceleration parameter calculated as
$q$=$(1-3k)/(k+1)$. From the deceleration parameter and
(\ref{scale1}), we see that the scale factor, $a(t)$, is up to the
parameter $k$, and it corresponds to these different cosmological
models for $k$ values: firstly, for $k\in
(-\infty,-1)\cup(1/3,\infty)$ the deceleration parameter becomes
$q<0$, and hence obey an accelerated power law expansion; secondly,
for $k\in(-1,1/3)$, the deceleration parameter becomes $q>0$, and
therefore a decelerated expansion occurs; thirdly, for the $k$=$1/3$
the deceleration parameter becomes $q$=$0$, and this case
corresponds to a uniformly (i.e. monotonically) expanding universe
model. Also, for $k$=$1/3$, we have
\begin{eqnarray}
a(t)=\frac{I_{1}}{4f_{0}\sqrt{\Psi_{0}}}t + a_0. \label{scl2}
\end{eqnarray}
This solution, corresponding to a universe with the pressureless
Dirac field, is similar to the matter (or dust) dominant universe in
the standard $2+1$ dimensional general relativity \cite{BarrowS06}.
Therefore, this solution shows that the Dirac field behaves as a
standard pressureless matter field in the $2+1$ dimensional
teleparallel gravity.

Furthermore, we can define the equation of state parameter of the
fermionic field by using the energy density (\ref{ener}) and
pressure (\ref{pressure}) as $\omega_{f}$=$p_{_f}/\rho_{_f}$ to
search that whether the fermionic field
can provide alternative for dark energy or not. Given the equations (%
\ref{coupling}), (\ref{pot1}), (\ref{sol2}) and (\ref{scale1}), we
obtain
\begin{eqnarray}
\omega_{_f} = \frac{1-3k}{k+1}.  \label{eos}
\end{eqnarray}
Based on the recently astrophysical observational data
\cite{Ade2013, Kumar-Xu2014}, the equation of the state parameter
tends to value $-1$. Additionally, as the equation of the state
parameter is less than $-1$ the dark energy is described by phantom,
but, for $\omega\in(-1, -1/3)$, the quintessence dark energy is
observed and the case $\omega$=$-1$ corresponds to the cosmological
constant. In the present work, we observed that, the case $k\in(1/2,
1)$ corresponding to the quintessence phase, and the case $k\in
(-\infty,-1)\cup(1,\infty)$ is corresponding to the phantom phase.
In both cases, the universe is both expanding and accelerating.
Therefore, the results show that the fermionic field may behave like
both the quintessence and the phantom dark energy field in the
late-time universe.

For the case $k$=$1$, the coupling and potential functions have
linear forms of $\Psi$ from the equation (\ref{coupling}) and
(\ref{pot1}) as follows:
\begin{eqnarray*}
F(\Psi) = f_{0}\Psi,  \label{coupling2}
\end{eqnarray*}
\begin{eqnarray}
V(\Psi) = \lambda\Psi.  \label{pot2}
\end{eqnarray}
Therefore, without the Noether symmetry, the Friedmann equation
(\ref{fried}) is reduced to
\begin{eqnarray}
\frac{\dot{a}}{a} -\sqrt{\frac{\lambda}{2 f_{0}}} =0,  \label{sol-4}
\end{eqnarray}
which has the following solution
\begin{eqnarray}
a(t) =a_0 e^{H_0 t}, \quad \mathrm{where} \quad
H_0=\sqrt{\frac{\lambda}{2 f_{0}}},  \label{sol-5}
\end{eqnarray}
where $a_0$ is a constant. It is clear that this solution is
corresponding to the de Sitter solution that describes an
inflationary epoch of the universe. Hence, it can be concluded that
the Dirac field plays a role as an inflaton field in the $2+1$
dimensional teleparallel gravity. In this situation, from Eqs.
(\ref{ener}) and (\ref{pressure}), the energy density and the
pressure of the Dirac field are given by
\begin{eqnarray}
\rho_{_f} = \frac{\lambda \Psi_0}{a_0^2}e^{-2 H_0 t}, \qquad
p_{_f}=-\rho_{_f}.  \label{sol-eden}
\end{eqnarray}
Furthermore, the state parameter calculated as $\omega_{_f}$=$-1$
corresponds to the cosmological constant.

It is interesting to note that in the case of vanishing the gauge
term (i.e., $B$=$0$), the partial differential equations
Eqs.(\ref{neq3})-(\ref{neq9}) are reduced to the equations for the
usual Noether symmetry approach. In this case, in a similar form of
the Eqs.(\ref{scale1}), the scale factor obtained as
\begin{eqnarray}
a\left( t\right)=\left[\widetilde{a}_{0}\left( k-1\right) \left(
t-a_{1}\right) \right]^{\frac{1}{1-k}} \label{scaleNS}
\end{eqnarray}
where $a_{1}$ integration constant and
$\widetilde{a}_{0}$=$\frac{\lambda \Psi_{0}^{1-k}}{2f_{0}}$. This
solution is the same as the results obtained in the context of
non-minimal coupling the fermionic field to the torsion in 3+1
dimensional teleparallel gravity \cite{Y}. According to this
solution, the deceleration parameter becomes $q$=$-k$, and hence,
the universe is accelerating for $k>0$, decelerating for $k<0$ and
the case $k<0$ is corresponds to uniformly expanding universe model.
Also, the equation of state parameter, $\omega$, calculated as
$\omega$=$-k$, so that, the cases $k\in(1/3, 1)$, $k\in(1,\infty)$
and $k$=$1$ are correspond to the quintessence phase, to the phantom
phase and to the cosmological constant, respectively. Hence, we can
say that the fermionic field causes the same physical effect on the
universe's history in both 2+1 and 3+1 dimensional teleparallel
gravity theories under the Noether symmetry condition. On the other
hand, there is a similar equivalence between the 2+1 \cite{GYY} and
3+1 \cite{Kremer} dimensional Einstein gravities under Noether
symmetry approach in the presence of the non-minimal coupling the
fermionic field to curvature scalar.

On the other hand, the minimal coupling case is characterized by
$F^{\prime}$=$0$, i.e. by constant coupling function $F$=$n$. If we
choose $n$=$1/2$ and ignore the gauge term (i.e., $B$=$0$), the
solutions of the partial differential equations (\ref{neq3}) are
\begin{eqnarray}
\alpha &=\alpha_{0}, \quad \beta_{j}=(\epsilon _{j}\beta
_{0}-\frac{\alpha_{0}}{a})\psi_{j},  \nonumber \\&
\gamma_{j}=(-\epsilon _{j}\beta _{0}-\frac{\alpha_{0}}{a})\psi
_{j}^{\dagger} \label{newvec1}
\end{eqnarray}
and from the Eqs.(\ref{neq9}) the self-interaction potential is
determined as
\begin{eqnarray}
V(\Psi) = V_{0}\Psi.  \label{newpot1}
\end{eqnarray}
Hence, from the (\ref{acce}) and (\ref{fried}), the scale factor is
calculated as,
\begin{eqnarray}
a(t)=\sqrt{V_{0}\Psi_{0}}t + a_0. \label{newscl2}
\end{eqnarray}

According to (\ref{newscl2}), the deceleration parameter becomes
$q=0$, Hence, we can say that, in the minimal coupling case, the
universe in the 2+1 dimensional teleparallel gravity in the presence
of the Dirac field is expanding uniformly, only. Furthermore, this
solution describes a universe dominated by pressureless matter (or
dust) \cite{BarrowS06}. Also, this solution is equivalent to the
solution of the 2+1 Einstein general relativity in the case of the
minimal coupling of fermionic fields with curvature scalar
\cite{GYY}. On the other hand, in the non-minimal coupling case, 2+1
dimensional teleparallel gravity has a similar physical results
about the history of the universe that of 3+1 dimensional
teleparallel gravity \cite{Y}, whereas it is different from the 3+1
dimensional standard Einstein theory, because, the 3+1 dimensional
standard Einstein theory explains the early-time inflation period,
only \cite{Kremer}, but, as shown in this study, 2+1 dimensional
teleparallel gravity explains both the early-time and late-time
inflation periods.

\section{Concluding remarks}

\label{conc}

The investigations to obtain a suitable components that can explain
the early-time inflation and late-time acceleration epoch of the
universe is one of the main topics of modern cosmology. In this
context, we consider the non-minimal coupling of the Dirac field to
torsion in the context of teleparallel theory of gravity in 2+1
dimensional Weitzenbock spacetime to investigate whether dirac
fields may be responsible for the early-time inflation and late-time
acceleration. Therefore, the identification of the non-minimal
coupling and the self-interaction functions are performed by using
the Noether symmetry approach with and without a gauge term. From
this approach, the obtained results can be summarized as follows:

\begin{itemize}
\item For the case $k$=$1$, the scale factor obtained in Eqs. (\ref{sol-5}) is $a(t)$=$a_0 e^{H_0 t}$, indicate the inflation epoch of the
universe. Also this solution corresponds to the vacuum energy which
characterized by cosmological constant.

\item The case $k\in(1/2, 1)$ conform to the quintessence dark energy
component that is used to explain the late-time acceleration.
\item The case $k\in(-\infty,-1)\cup(1,\infty)$ corresponds to the phantom dark
energy component. This component is, also, used to explain the
late-time acceleration.
\item The cosmological solution for $k$=$1/3$ corresponds to a uniformly (i.e. monotonically,) expanding universe model. Also, it is
obtained a solution correspond to the matter (or dust) dominant
universe with the pressure of the Dirac field $p_{_f}$=$0$ as in the
$2+1$ dimensional General Relativity \cite{BarrowS06}. Hence, it
shows that the Dirac field behaves as a standard pressureless matter
field in the $2+1$ dimensional teleparallel gravity.
\item Under the usual Noether symmetry condition (i.e., $B$=$0$), the scale factor $a(t)$ is computed as $a\left(
t\right)$=$\left[\widetilde{a}_{0}\left( k-1\right) \left(
t-a_{1}\right) \right]^{\frac{1}{1-k}}$, where $a_{1}$ integration
constant and $\widetilde{a}_{0}$=$\frac{\lambda
\Psi_{0}^{1-k}}{2f_{0}}$. It is interesting to note that this
solution is the same as the results obtained in the context of
non-minimal coupling the fermionic field to the torsion in 3+1
dimensional teleparallel gravity \cite{Y}. Hence, we can say that,
in the presence of the fermionic field, the 2+1 dimensional
teleparallel gravity has similar physical results about the
evolution of the universe that of the 3+1 dimensional teleparallel
gravity thanks to the Noether symmetry.
\item In the case of the minimal coupling (i.e., $F$=$1/2$) and under the usual Noether symmetry condition (i.e., $B$=$0$)
our solutions are reduced to 2+1 dimensional Einstein general
relativity solutions, as well.
\item It is important to emphasize that for the non-minimal coupling and under the Noether symmetry
approach with a gauge term, the 2+1 dimensional Einstein gravity
\cite{GYY} has similar physical results about the history of the
universe that of both 2+1 and 3+1 dimensional teleparallel gravities
\cite{Y} in the presence of the fermionic field.
\end{itemize}

Finally, it can be said that the Dirac field is an inflaton
describing the acceleration of the Universe both in the early time
epoch and in the late-time epoch.

\section*{Acknowledgments}

The Authors thanks Dr. Yusuf Kucukakca for useful discussion. Also,
the authors are grateful to the anonymous referees for their
valuable comments. This work was supported by the Scientific
Research Projects Unit of Akdeniz University.


\end{document}